\documentclass[12pt]{article}%
\usepackage{amsmath}
\usepackage{amsfonts}
\usepackage{amssymb}
\usepackage{graphicx}%
\providecommand{\U}[1]{\protect\rule{.1in}{.1in}}
\newcommand{\bfr}{\begin{flushright}}
\newcommand{\efr}{\end{flushright}}
\newcommand{\bc}{\begin{center}}
\newcommand{\ec}{\end{center}}
\newcommand{\ben}{\begin{enumerate}}
\newcommand{\een}{\end{enumerate}}

\newcommand{\be}{\begin{equation}}
\newcommand{\ee}{\end{equation}}
\newcommand{\ba}{\begin{array}}
\newcommand{\ea}{\end{array}}
\def\6{\partial}
\begin{document}

\title{\textbf{Plane waves in noncommutative fluids}}
\author{M. C. B. Abdalla$^{a}$\thanks{email: mabdalla@ift.unesp.br},\thinspace\ L.
Holender$^{b}$\thanks{email: holender@ufrrj.br},\thinspace\ M. A. Santos$^{c}%
$\thanks{email: masantos@cce.ufes.br}\thinspace\ and I. V. Vancea$^{b}%
$\thanks{email: ionvancea@ufrrj.br}\\$^{a}$\emph{{\small Instituto de F{\'{\i}}sica Te\'{o}rica, UNESP -
Universidade Estadual Paulista,}}\\\emph{{\small Rua Dr. Bento Teobaldo Ferraz 271, Bloco 2, Barra-Funda,}}\\\emph{{\small Caixa Postal 70532-2, 01156-970, S\~ao Paulo - SP, Brasil}}\\$^{b}$\emph{{\small Grupo de F{\'{\i}}sica Te\'{o}rica e Matem\'{a}tica
F\'{\i}sica, Departamento de F\'{\i}sica,}}\\\emph{{\small Universidade Federal Rural do Rio de Janeiro (UFRRJ),}}\\\emph{{\small Cx. Postal 23851, BR 465 Km 7, 23890-000 Serop\'{e}dica - RJ,
Brasil }}\\$^{c}$\emph{{\small Departamento de F\'{\i}sica e Qu\'{\i}mica,}}\\\emph{{\small Universidade Federal do Esp\'{\i}rito Santo (UFES),}}\\\emph{{\small Avenida Fernando Ferarri S/N - Goiabeiras, 29060-900 Vit\'{o}ria
- ES, Brasil}}}
\date{03 March 2013}
\maketitle

\thispagestyle{empty}


\abstract{We study the dynamics of the noncommutative fluid in the Snyder space perturbatively at the first order in powers of the noncommutative parameter.
The linearized noncommutative fluid dynamics is described by a system of coupled linear partial differential equations in which the variables are the fluid density and the fluid potentials.
We show that these equations admit a set of solutions that are monocromatic plane waves for the fluid density and two of the potentials and a linear function for the third potential.
The energy-momentum tensor of the plane waves is calculated.}

\vfill


\newpage\pagestyle{plain} \pagenumbering{arabic}

\section{Introduction}

Motivated by recent works in which several noncommutative systems of charges
were shown to display fluid properties at quantum scale
\cite{Bahcall:1991an,Susskind:2001fb,ElRhalami:2001xf,Barbon:2001dw,Barbon:2007zz,Polychronakos:2007df,
Alavi:2006sr} as well as cosmological scale \cite{DeFelice:2009bx}, we have
proposed the first field theoretical model of fluid in the canonical
noncommutative space \cite{Holender:2011px} and the noncommutative Lorentz
covariant space \cite{Abdalla:2012tt}, respectively
\cite{Snyder:1946qz,Snyder:1947nq, AmelinoCamelia:2001fd}. The noncommutative
fluid was constructed as a noncommutative field theory
\cite{Douglas:2001ba,Szabo:2001kg} in the realization method approach
\cite{Jonke:2001xk,Meljanac:2006ui,KresicJuric:2007nh,Meljanac:2007xb,Govindarajan:2008qa,Battisti:2008xy,Govindarajan:2009wt,Meljanac:2009fy,Battisti:2010sr,Meljanac:2011mt,Lukierski:1993wx,Banerjee:2006wf,Ghosh:2007ai}
by generalizing the first order action functional of the commutative perfect
relativistic fluid
\cite{Jackiw:2002pn,Jackiw:2002tw,Jackiw:2003dw,Jackiw:2004nm}. In this
approach one obtains the fluid dynamics of the long wavelength degrees of
freedom of the system bypassing the statistical analysis of the microscopic
degrees of freedom which in the noncommutative spaces is not well understood
yet (see for tentative approaches
\cite{Alavi:2007dr,Huang:2009yx,Marcial:2010zza}). In general, the definition
of the noncommutative fluid degrees of freedom is not a trivial task since the
phenomenological considerations that led to the fluid equations in the
commutative spaces are not allowed. However, by treating the perfect fluid as
an effective field theory, the degrees of freedom of the noncommutative fluid
are defined by the correspondence principle that requires that in the
commutative limit of the noncommutative parameter the known commutative
perfect fluid equations be obtained. As was discussed in
\cite{Holender:2011px, Abdalla:2012tt}, the degrees of freedom of the
noncommutative fluid are the noncommutative generalization of the density
current and the fluid potentials that parametrize the velocity field
\cite{Jackiw:2000mm,Carter-lectures}. The choice of the commutative fluid
potentials is not unique. When it is made in terms of real functions
$\theta(x)$, $\alpha(x)$ and $\beta(x)$ it is called the Clebsch
parametrization \cite{Jackiw:2000mm,Carter-lectures} while the fluid
potentials given in terms of one real $\theta(x)$ and two complex functions
$z(x)$ and $\bar{z}(x)$, respectively, define the so called K\"{a}hler
parametrization
\cite{Nyawelo:2003bv,Jackiw:2000cc,Baleanu:2004sc,Jarvis:2005hp,Nyawelo:2003bw,Grassi:2011wt,Holender:2008qj,Holender:2012uq}%
.

The noncommutative coordinates in the Snyder space $\mathcal{S}$ are the Lie
generators of $so(1,4)/so(1,3)$. The degrees of freedom of the noncommutative
fluid in $\mathcal{S}$ defined in \cite{Abdalla:2012tt} are the natural
generalization of the commutative fluid potentials in the Clebsch
parametrization. These belong to the set of functions over the Snyder space
$\mathcal{F}(\mathcal{S})$ which can be endowed with the star-product and the
co-product constructed in \cite{Girelli:2009ii,Girelli:2010wi}. The algebra
$\mathcal{F}(\mathcal{S})$ is isomorphic to the deformed algebra over the
Minkowski space-time $\left(  C^{\infty}(\mathcal{M}),\star\right)  $. Since
the star-product is nonassociative and noncommutative and the momenta
associated to the coordinates do not form a Lie group, understanding the
dynamics of the noncommutative fluid in the Snyder space turns to be a very
challenging problem. The reason is that the noncommutative fluid equations
involve an infinite number of derivatives of the fluid potentials which are
multiplied in a nonassociative way. This property makes the equations
difficult to analyse in the general case. However, if the deformation
parameter of the deformed Poincar\'{e} algebra $s=l_{s}^{2}$, where $l_{s}$ is
the typical length scale of the noncommutative space is small compared to
unity, one can attempt to expand the star-product in powers of $s$. This is
certainly the case if the noncommutative structure is the structure of the
physical space-time since the phenomenological data and the theoretical
arguments suggest that $l_{s}$ is of the order of the Planck scale. Another
reason for which one should study the terms of the action (\ref{action-star})
corresponding to the finite order in $s$ is the following. Due to the
nonassociativity of the star-product (\ref{star-prod-1}), the highest order of
the derivatives of the fluid potentials from the noncommutative functional is
infinite. Therefore, the Euler-Lagrange equations of motion cannot be
determined for an arbitrary value of $s$. Moreover, since the theory has just
a limited number of symmetries, the equations of motion are not integrable.

In this paper, we are going to study perturbatively the dynamics of the
noncommutative fluid in the Snyder space by expanding the relevant objects
from the algebra $\left(  C^{\infty}(\mathcal{M}),\star\right)  $ in powers of
$s$. Our main goal is to obtain analytic solutions to the equations of motion
of the fluid density current and potentials, respectively, in the linear
approximation. The dependence on $s$ is encoded in the function on two momenta
that determines the star-product and the co-product of the deformed
Poincar\'{e} algebra and the anti-pode of its co-algebra. By truncating the
power expansion at first order in $s$, the noncommutative fluid equations are
reduced to a system of coupled linear partial differential equations. We are
able to show that these equations admit monocromatic plane wave solutions for
the density current $j^{\mu}$ and $\alpha$ and $\beta$ fluid potentials,
respectively and a linear solution for $\theta$ potential. These are the first
solutions of the linearized noncommutative fluid dynamics obtained so far.
Also, we calculate the energy-momentum tensor of the monocromatic waves
explicitely, which represents one of the few examples of such calculations.

The paper is organized as follows. In Section 2, we give a short review of the
recent noncommutative fluid model in the Snyder space obtained in
\cite{Abdalla:2012tt} and establish our notations. The perturbative expansion
of the model and its first order equations of motion are obtained in Section
3. In Section 4, we show that the equations of motion admit solutions that are
monocromatic plane waves of the fluid potentials. These solutions are
characterized by the fact that the scalar product of divergence of the
potentials $\alpha$ and $\beta$ with the current density $j^{\mu}$ in the
Minkowski space-time is zero. Also, we calculate the energy-momentum tensor
for these solutions. In the last section we discuss the properties of the
solutions obtained previously.

\section{Noncommutative fluid in the Snyder space}

The Snyder-like geometries can be viewed as realizations of the Snyder algebra
which, at its turn, is a deformation of the algebra $so(1,3)$ with the
deformation parameter $s=l_{s}^{2}$
\begin{align}
\left[  \tilde{x}_{\mu},\tilde{x}_{\nu}\right]   &  =sM_{\mu\nu}%
,\label{alg-Snyder-x}\\
\left[  p_{\mu},p_{\nu}\right]   &  =0,\label{alg-Snyder-p}\\
\left[  M_{\mu\nu},M_{\rho\sigma}\right]   &  =\eta_{\nu\rho}M_{\mu\sigma
}-\eta_{\mu\rho}M_{\nu\sigma}+\eta_{\mu\sigma}M_{\nu\rho}-\eta_{\nu\sigma
}M_{\mu\rho},\label{alg-Snyder-M}\\
\left[  M_{\mu\nu},\tilde{x}_{\rho}\right]   &  =\eta_{\nu\rho}\tilde{x}_{\mu
}-\eta_{\mu\rho}\tilde{x}_{\nu},\label{alg-Snyder-Mx}\\
\left[  M_{\mu\nu},p_{\rho}\right]   &  =\eta_{\nu\rho}p_{\mu}-\eta_{\mu\rho
}p_{\nu}, \label{alg-Snyder-Mp}%
\end{align}
where $\mu,\nu=\overline{0,3}$ and $l_{s}$ has the dimension of length. The
generators $M_{\mu\nu}$ can be expressed in terms of the commutative
coordinates and momenta $\{x_{\mu},p_{\nu}\}$ of the Minkowski space-time
$\mathcal{M}$ in the usual fashion $M_{\mu\nu}=i(x_{\mu}p_{\nu}-x_{\nu}p_{\mu
})$. The noncommutative spaces compatible with the Lorentz symmetry can be
obtained by realizing geometrically the Snyder algebra (\ref{alg-Snyder-x}%
)-(\ref{alg-Snyder-Mp}) through the so called realization method. In these
spaces, the generators $\tilde{x}_{\mu}$ are interpreted as noncommutative
position operators associated to the sites of a lattice of typical length
$l_{s}$. The closure of the commutators over $so(1,3)$ is equivalent to the
statement that the lattice space is compatible with the Lorentz symmetry
\cite{Snyder:1946qz}. The functions $\tilde{x}_{\mu}(x,p)$ and their
commutation relations with the generators $p_{\mu}$ are not determined by the
Snyder algebra \cite{Battisti:2008xy}. The realization method allows one to
construct noncommutative spaces in which the functions $\tilde{x}_{\mu}(x,p)$
are momentum dependent rescalings of the coordinates
\begin{equation}
\tilde{x}_{\mu}(x,p)=\Phi_{\mu\nu}(s;p)x_{\nu}. \label{representations-1}%
\end{equation}
It can be shown that the smooth functions take the form $\Phi_{\mu\nu
}(s;p)=\Phi_{\mu\nu}\left[  \varphi(s;p)\right]  $ such that%
\begin{equation}
\tilde{x}_{\mu}(x,p)=x_{\mu}\varphi(A)+s\left\langle xp\right\rangle p_{\mu
}\left[  1+2\frac{d\varphi(A)}{dA}\right]  \left[  \varphi(A)-2A\frac
{d\varphi(A)}{dA}\right]  ^{-1}, \label{rep-2}%
\end{equation}
where $A=s\eta^{\mu\nu}p_{\mu}p_{\nu}$ and the commutative scalar product is
denoted by $\left\langle ab\right\rangle =\eta^{\mu\nu}a_{\mu}b_{\nu}$. From
the above equations, one can see that the Snyder geometry is a non-canonical
deformation of the commutative phase space of coordinates $\{x_{\mu},p_{\mu
}\}$. The realization formalism allows one to work simultaneously with various
noncommutative spaces which are characterized by different functions
$\varphi(A)$. For example, the Weyl, the Snyder and the Maggiore
noncommutative space-times can be obtaining by choosing $\varphi(A)=\sqrt
{A}\cot(A)$, $\varphi(A)=1$ and $\varphi(A)=\sqrt{1-sp^{2}}$, respectively.
Interpolations among these spaces are also possible \cite{Battisti:2010sr}.

The noncommutative fluid constructed in \cite{Abdalla:2012tt} generalizes the
perfect relativistic fluid models in the Clebsch parametrization. The dynamics
of the commutative fluid can be derived from an action functional $S[\phi(x)]$
that depends on the density current and three real fluid potentials
$\phi(x)=\{j^{\mu}(x),\theta(x),\alpha(x),\beta(x)\}$ by applying field
theoretical methods \cite{Jackiw:2000mm}. The action of the noncommutative
fluid is determined by a correspondence principle that constraints the
possible noncommutative functionals such that the equations of the
relativistic fluid are obtained in the commutative limit
\begin{equation}
\lim_{s\rightarrow0}S_{s}[\tilde{\phi}(\tilde{x})]=S[\phi
(x)].\label{cor-princ}%
\end{equation}
It follows that the fluid potentials must be generalized to the functions
$\tilde{\phi}(\tilde{x})=\{\tilde{j}^{\mu}(\tilde{x}),\tilde{\theta}(\tilde
{x}),\tilde{\alpha}(\tilde{x}),\tilde{\beta}(\tilde{x})\}$ from $\mathcal{F}%
(\mathcal{S})$ that should be identified with the degrees of freedom of the
noncommutative fluid. It is useful to map the noncommutative and
nonassociative algebra $\mathcal{F}(\mathcal{S})$ into the deformed algebra of
the Minkowski space-time $\left(  C^{\infty}(\mathcal{M}),\star\right)  $ as
follows. If $\tilde{\phi}(\tilde{x})$ is a noncommutative function and
$\mathbf{1}$ is the identity element of the algebra of commutative functions
over $x_{\mu}$ then%
\begin{equation}
\tilde{\phi}(\tilde{x})\vartriangleright\mathbf{1}=\psi(x),\label{co-alg-def}%
\end{equation}
where $\psi(x)$, in general, differs from $\tilde{\phi}(x)$. The star-product,
the co-product and the anti-pode $S$ of the Poincar\'{e} co-algebra are
defined by the following relations%
\begin{align}
e^{i\left\langle K_{1}^{-1}(k_{1})\tilde{x}\right\rangle }\star
e^{i\left\langle K_{2}^{-1}(k_{2})\tilde{x}\right\rangle } &
=e^{i\left\langle D^{(2)}(k_{2},k_{1})x\right\rangle },\label{star-prod}\\
\triangle p_{\mu} &  =D_{\mu}^{(2)}(p\otimes\mathbf{1},\mathbf{1}\otimes
p),\label{co-prod}\\
D_{\mu}^{(2)}\left(  g,S(g)\right)   &  =0,\label{anti-pode}%
\end{align}
for any element $g$ of the deformed Poincar\'{e} group which has its
co-algebra structure deformed according to the equation (\ref{co-prod}) while
its Poincar\'{e} algebra is undeformed as given by the equations
(\ref{alg-Snyder-x}) - (\ref{alg-Snyder-Mp}). The non-commutative functions
$e^{i\left\langle k\tilde{x}\right\rangle }$ depend on the deformed momentum
$K_{\mu}=K_{\mu}(k)$ and are defined by the following relation%
\begin{equation}
e^{i\left\langle k\tilde{x}\right\rangle }\vartriangleright\mathbf{1=}%
e^{i\left\langle K\tilde{x}\right\rangle }.\label{k-momentum}%
\end{equation}
It follows from the equations (\ref{star-prod}) - (\ref{anti-pode}) that the
two-functions $D^{(2)}(k_{2},k_{1})$ determine completely the algebraic
structure of the deformed algebra. By choosing the differential representation
of the generators $p_{\mu}=-i\partial_{\mu}$ the star-product can be written
as \cite{Meljanac:2007xb}%
\begin{equation}
\left(  f\star g\right)  (x)=\lim_{y\rightarrow x}\lim_{z\rightarrow x}%
\exp\left[  i\left\langle \left(  D^{(2)}(p_{y},p_{z})-p_{y}-p_{z}\right)
x\right\rangle \right]  .\label{star-prod-1}%
\end{equation}
As was shown in \cite{Abdalla:2012tt} the action of the noncommutative fluid
that obeys the correspondence principle (\ref{cor-princ}) has the following
form%
\begin{align}
S_{s} &  \left[  j^{\mu}(x),\theta(x),\alpha(x),\beta(x)\right]  =%
{\displaystyle\int}
d^{4}x\tilde{\mathcal{L}}[\tilde{\theta}(\tilde{x}),\tilde{\alpha}(\tilde
{x}),\tilde{\beta}(\tilde{x})]\vartriangleright\mathbf{1}\nonumber\\
&  =%
{\displaystyle\int}
d^{4}x\left[  -j^{\mu}(x)\star\left[  \partial_{\mu}\theta(x)+\alpha
(x)\star\partial_{\mu}\beta(x)\right]  -f_{s}\left(  \sqrt{-j^{\mu}(x)\star
j_{\mu}(x)}\right)  \right]  ,\label{action-star}%
\end{align}
where the last equality defines a functional over $\left(  C^{\infty
}(\mathcal{M}),\star\right)  $. The function $f_{s}$ from $\left(  C^{\infty
}(\mathcal{M}),\star\right)  $ is the image under the map (\ref{co-alg-def})
of an arbitrary function $\tilde{f}$ from $\mathcal{F}(\mathcal{S})$. It
follows that the action (\ref{action-star}) describes a class of
noncommutative fluids parametrized by $f_{s}$ for any given value of $s$.

In general, due to the lack of phenomenological information about the
noncommutative fluid, it is difficult to define the relevant physical
quantities such as the energy and momentum densities of the fluid. The
functional approach to the noncommutative fluid has the advantage of making
the definition of these quantities conceptually simpler, although their
computation is difficultated by the nonassociativity of the star-product. The
energy and momentum can be defined by the variation of the action
(\ref{action-star}) under infinitesimal deformed translations $\delta
_{\varepsilon}x_{\mu}$ that can be obtained from the deformed Poincar\'{e}
transformations. It was shown in \cite{Abdalla:2012tt} that the variation of
the action under the deformed translations results in the following
energy-momentum tensor%
\begin{equation}
T_{\nu}^{\mu}=\Theta_{\sigma}^{\mu}(\phi)-\mathcal{L}\eta_{\nu}^{\mu},
\label{A-2}%
\end{equation}
where $\Theta^{\mu\nu}(\phi)$ is a functional of the fluid potentials and
their derivatives up to the third order. An outstandingly difficult problem
created by the noncommutativity and the nonassociativity of the star-product
is to determine and to solve the equations of motion which, in general, form a
system of nonlinear partial differential equations of arbitrarily high order
and to obtain an analytic formula for $T_{\nu}^{\mu}$. This task becomes
tractable at finite order in the powers of the noncommutative parameter $s$.

\section{Lower order expansion in $s$}

In this section, we are going to determine the equations of motion and the
variation of the action under the deformed Poincar\'{e} transformations at the
first order in $s$.

\subsection{First order action and equations of motion}

The nonassociative exponential from the star-product contains infinitely many
derivatives of the fluid potentials. Therefore, the truncation of the
star-product to some finite order in the powers of $s$ is needed. The $s$
dependence of the $\star$-product is encoded in the two-functions $D_{\mu
}^{(2)}(k_{1},k_{2})$ \cite{KresicJuric:2007nh}%
\begin{equation}
D_{\mu}^{(2)}(k_{1},k_{2})=\sum\limits_{n=1}^{\infty}s^{n}D_{\mu}^{(2)n}%
(k_{1},k_{2}).\label{power-D}%
\end{equation}
In order to obtain the linearized action in $s$, we consider only the first
two terms from the above series%
\begin{align}
D_{\mu}^{(2)0}(k_{1},k_{2}) &  =k_{1},_{\mu}+k_{2},_{\mu},\label{D-0}\\
D_{\mu}^{(2)1}(k_{1},k_{2}) &  =A(k_{1},k_{2})k_{1},_{\mu}+B(k_{1},k_{2}%
)k_{2},_{\mu},\label{D-1}%
\end{align}
where the functions $A(k_{1},k_{2})$ and $B(k_{1},k_{2})$ have the following
form%
\begin{align}
A(k_{1},k_{2}) &  =c\left(  k_{2}^{2}+2k_{1}k_{2}\right)  ,\label{a-b-const-1}%
\\
B(k_{1},k_{2}) &  =\left(  c-\frac{1}{2}\right)  k_{1}^{2}+\left(  2c-\frac
{1}{2}\right)  k_{1}k_{2},\label{a-b-const-2}\\
c &  =\frac{2c_{1}+1}{2}.\label{a-b-const-3}%
\end{align}
The real constant $c_{1}$ is realization dependent and has the following
values: $c_{1}=-1/2$ for the Maggiore, $c_{1}=0$ for the Snyder and
$c_{1}=-1/3$ for the Weyl realizations, respectively. The first order action
can be obtained by linearizing simultaneously the star-product and the
two-functions with respect to $s$. Some algebra shows that the linearized
action takes the following form%
\begin{align}
S_{s} &  =-\int d^{4}x\left\{  j^{\mu}(x)\partial_{\mu}\theta(x)+ix^{\mu
}\left[  K_{\mu}^{s}(y,z)j^{\nu}(y)\partial_{\nu}\theta(z)\right]
|_{y=z=x}\right\}  \nonumber\\
&  -\int d^{4}x\left\{  j^{\mu}(x)\alpha(x)\partial_{\mu}\beta(x)+ix^{\mu
}\left[  K_{\mu}^{s}(w,x)j^{\nu}(w)\alpha(x)\partial_{\nu}\beta(x)\right]
|_{w=x}\right\}  \nonumber\\
&  -i\int d^{4}xj^{\nu}(x)x^{\mu}\left[  K_{\mu}^{s}(y,z)j^{\nu}%
(y)\partial_{\nu}\theta(z)\right]  \alpha(y)\partial_{\nu}\beta(z)|_{y=z=x}%
\nonumber\\
&  +\int d^{4}xx^{\mu}x^{\rho}K_{\mu}^{s}(w,x)K_{\rho}^{s}(y,z)j^{\nu
}(w)\alpha(y)\partial_{\nu}\beta(z)|_{w=y=z=x}\nonumber\\
&  +s\int d^{4}xx^{\mu}x^{\rho}\left[  D_{\mu}^{(2)0}(-i\partial_{\mu}%
^{w},-i\partial_{\mu}^{z})+i\partial_{\mu}^{w}+i\partial_{\mu}^{z}\right]
D_{\mu}^{(2)1}(-i\partial_{\mu}^{y},-i\partial_{\mu}^{z})j^{\nu}%
(w)\alpha(y)\partial_{\nu}\beta(z)|_{w=y=z=x}\nonumber\\
&  +\int d^{4}xf\left(  -\left(  1+ix^{\mu}\left[  K_{\mu}^{s}(y,z)j^{\nu
}(y)j_{\nu}(z)\right]  |_{y=z=x}\right)  ^{1/2}\right)  ,\label{S-lin}%
\end{align}
where we have used the momentum representation $k_{\mu}=-i\partial_{\mu}$ and
the following notation%
\begin{equation}
K_{\mu}^{s}(y,z)=D_{\mu}^{(2)0}(-i\partial_{\mu}^{y},-i\partial_{\mu}%
^{z})+sD_{\mu}^{(2)1}(-i\partial_{\mu}^{y},-i\partial_{\mu}^{z})+i\partial
_{\mu}^{y}+i\partial_{\mu}^{z}.\label{K-not}%
\end{equation}
By direct and lenghty calculations one can show that the equation of motion of
$j^{\mu}$ has the following form%
\begin{align}
&  -\partial_{\mu}\theta-\alpha\partial_{\mu}\beta+\partial_{\nu}\left(
\frac{f^{\prime}}{\rho_{0}}\right)  x^{\nu}j_{\mu}+\frac{5}{2}s\left[
\partial^{2}\partial_{\mu}\theta+\partial^{2}\left(  \alpha\partial_{\mu}%
\beta\right)  \right]  \nonumber\\
&  +\frac{1}{2}sx^{\nu}\left[  \partial^{2}\alpha\partial_{\nu}\partial_{\mu
}\beta+2\partial_{\rho}\alpha\partial^{\rho}\partial_{\nu}\partial_{\mu}%
\beta+\partial_{\rho}\partial_{\nu}\alpha\partial^{\rho}\partial_{\mu}%
\beta\right]  \nonumber\\
&  +is\partial_{\nu}\left[  \frac{f^{\prime}}{2\rho_{0}}\left[  \left(
2c-\frac{1}{2}\right)  \delta^{\nu\sigma}x_{\sigma}\partial^{2}j_{\mu}+\left(
4c-\frac{1}{2}\right)  x_{\sigma}\partial^{\sigma}\partial^{\nu}j_{\mu
}\right]  \right]  \nonumber\\
&  +is\partial_{\nu\omega}^{2}\left[  \frac{f^{\prime}}{2\rho_{0}}x_{\sigma
}\left[  \left(  2c-\frac{1}{2}\right)  \eta^{\omega\nu}\partial^{\sigma
}j_{\mu}+\left(  4c-\frac{1}{2}\right)  \delta^{\omega\sigma}\partial^{\nu
}j_{\mu}\right]  \right]  =0.\label{eq-mot-j}%
\end{align}
Here, $f^{\prime}$ represents the derivative of $f$ with respect to its
argument $\rho_{0}=\sqrt{-j^{\mu}(x)\star j_{\mu}(x)}$. In a similar way, one
can derive the equation of motion for the fluid potential $\theta$%
\begin{equation}
\partial_{\mu}j^{\mu}+\frac{1}{2}s\partial^{2}\partial_{\mu}j^{\mu
}=0.\label{eq-mot-theta-1}%
\end{equation}
The equation of motion of the potential $\alpha$ is given by the following
relation%
\begin{align}
&  -j^{\mu}\partial_{\mu}\beta+s\left(  \partial^{2}j^{\mu}\partial_{\mu}%
\beta+2\partial_{\mu}j^{\nu}\partial^{\mu}\partial_{\nu}\beta+j^{\mu}%
\partial^{2}\partial_{\mu}\beta\right)  \nonumber\\
&  +\frac{s}{2}x^{\mu}\left(  \partial^{2}j^{\nu}\partial_{\mu}\partial_{\nu
}\beta+\partial_{\mu}j^{\nu}\partial^{2}\partial_{\nu}\beta+\partial_{\mu
}\partial_{\nu}j^{\sigma}\partial^{\nu}\partial_{\sigma}\beta+\partial_{\nu
}j^{\sigma}\partial_{\mu}\partial^{\nu}\partial_{\sigma}\beta\right)
=0.\label{eq-mot-alpha-1}%
\end{align}
Finally, one can show the equation of motion of the potential $\beta$ takes
the following form%
\begin{equation}
\partial_{\mu}\left(  j^{\mu}\alpha\right)  +s\left(  3\partial_{\mu}j^{\mu
}\partial^{2}\alpha+4\partial_{\mu}\partial_{\nu}j^{\mu}\partial^{\mu}%
\alpha+\frac{5}{2}\partial_{\mu}j^{\nu}\partial_{\nu}\partial^{\mu}%
\alpha+\partial^{2}j^{\mu}\partial_{\mu}\alpha+\partial^{2}\partial_{\mu
}j^{\mu}\alpha+j^{\mu}\partial^{2}\partial_{\mu}\alpha\right)
=0.\label{eq-mot-beta}%
\end{equation}
We note that the zeroth order terms of the equation of motion (\ref{eq-mot-j})
differ from the corresponding ones from the commutative equation of the
current $j^{\mu}$ by a term proportional to $x^{\nu}j_{\mu}$ which can be
cancelled by choosing proper boundary conditions. The equation
(\ref{eq-mot-theta-1}) shows that the current density is not conserved in the
noncommutative case which is a consequence of a lack of translation symmetry.
The violation of the translation invariance has been previously analysed in
\cite{Bertolami:2003nm}. Nevertheless, the conservation is restored in the
commutative limit. The lowest order expansion of the energy-momentum tensor
can be computed by expanding $T_{\nu}^{\mu}$ from the equation (\ref{A-2}) in
powers of $s$.

\section{Plane waves in noncommutative fluids in Snyder space}

The equations of motion (\ref{eq-mot-j}), (\ref{eq-mot-theta-1}),
(\ref{eq-mot-alpha-1}) and (\ref{eq-mot-beta}) obtained in the previous
section describe the dynamics of the noncommutative fluid at first order in
$s$. Therefore, in order to understand the dynamics of the noncommutative
fluid, it is important to look for analytic solutions to the system
(\ref{eq-mot-j}) - (\ref{eq-mot-beta}). In this section, we are going to
determine a particular set of solutions that are monocromatic waves of fluid
potentials. To this end, we observe that the equations (\ref{eq-mot-theta-1}),
(\ref{eq-mot-alpha-1}) and (\ref{eq-mot-beta}) form a subsystem of coupled
linear equations in each argument that involve the fluid potentials $j^{\mu
}(x)$, $\alpha(x)$ and $\beta(x)$ which belong to the algebra $\left(
C^{\infty}(\mathcal{M}),\left\langle \cdot\right\rangle \right)  $ with the
usual commutative product.

Let us solve the system by starting with the equation (\ref{eq-mot-theta-1})
which is independent of the potentials $\alpha(x)$ and $\beta(x)$ and is
second order and linear. By a simple field redefinition which lowers the
degree of the equation by one and for $s\neq0$ the equation
(\ref{eq-mot-theta-1}) is equivalent with the Klein-Gordon equation with the
mass parameter%
\begin{equation}
m_{\phi}^{2}=\frac{2}{s}.\label{KG-mass-1}%
\end{equation}
It follows that the equation of motion of the current density admits plane
wave solutions independent of the other fluid potentials%
\begin{equation}
j_{\pm}^{\mu}(x)=\mp i\mathbf{A}_{\pm}\frac{k^{\mu}}{k^{2}}e^{\pm
i\left\langle kx\right\rangle },\label{Sol-j-1}%
\end{equation}
where $\mathbf{A}_{\pm}$ is an arbitrary constant that determines the boundary
value of the density current and $k^{2}=\left\langle kk\right\rangle $. As can
be seen from the equation (\ref{eq-mot-alpha-1}), the solutions $j_{\pm}^{\mu
}(x)$ enter as coefficients in the equation of motion of the potential
$\beta(x)$. One can also lower the degree of (\ref{eq-mot-alpha-1}) by one and
after somewhat lengthy but straightforward calculations, one can show that
(\ref{eq-mot-alpha-1}) has plane wave solutions $\beta^{\pm}(x)$ which have
the property that the scalar products between their gradients with the current
density in the Minkowski space-time is zero%
\begin{equation}
\left\langle j(x)\partial\beta^{\pm}(x)\right\rangle =0.\label{Ort-j-beta}%
\end{equation}
The explicit form of the $\beta^{\pm}$ - waves is given by the following
equation%
\begin{equation}
\beta^{\pm}(x)=\mathbf{B}^{\pm}e^{\mp i\left\langle kx\right\rangle
},\label{Sol-beta-1}%
\end{equation}
where $\mathbf{B}^{\pm}$ are arbitrary constants specified by the boundary
conditions for the potential $\beta^{\pm}(x)$. The $\beta^{\pm}$ - waves move
in opposed direction to $j_{\pm}^{\mu}$ - waves. By following the same steps
as above, it is possible to show that the equation (\ref{eq-mot-beta}) has
plane wave solutions of gradients that satisfy the same condition%
\begin{equation}
\left\langle j(x)\partial\alpha^{\pm}(x)\right\rangle =0.\label{Ort-j-alpha}%
\end{equation}
The $\alpha^{\pm}$ - wave solutions have the following form%
\begin{equation}
\alpha^{\pm}(x)=\mathbf{C}^{\pm}e^{\pm i\left\langle kx\right\rangle
},\label{Sol-alpha-1}%
\end{equation}
where $\mathbf{C}^{\pm}$ are arbitrary constants determined by the boundary
value of the field $\alpha^{\pm}(x)$.

Let us focus now on the equation (\ref{eq-mot-j}) that describes the dynamics
of the potential $\theta$ as a function of the current density and the other
two fluid potentials. In order to find its solution, one needs to plug the
previous wave solutions (\ref{Sol-j-1}), (\ref{Sol-beta-1}) and
(\ref{Sol-alpha-1}) into (\ref{eq-mot-j}). After doing that and after some
more algebraic manipulations, we obtain the following equation%
\begin{align}
\left(  \partial^{2}-\frac{2}{5s}\right)  \partial_{\mu}\theta^{\pm}  &
=\pm\frac{2i\mathbf{A}_{\pm}}{5s}\frac{k_{\mu}}{k^{2}}x^{\nu}\partial_{\nu
}\left(  \frac{f^{\prime}}{\rho_{0}}\right)  e^{\pm i\left\langle
kx\right\rangle }\nonumber\\
&  -\frac{2i}{5}\left[  \left(  \Omega_{\mu}^{\pm}(k)\eta_{\nu}^{\sigma
}+\Lambda_{\pm\mu}^{\sigma\nu}(k)\right)  \partial_{\sigma}+\left(
\Gamma_{\pm\mu}^{\omega\sigma\nu}(k)+\Upsilon_{\pm\mu}^{\omega\sigma\nu
}(k)\right)  \partial_{\sigma\omega}^{2}\right]  \left(  \frac{f^{\prime}%
}{\rho_{0}}x^{\nu}e^{\pm i\left\langle kx\right\rangle }\right) \nonumber\\
&  -\frac{2i}{5s}\mathbf{D}_{\pm}k_{\mu}. \label{Sol-theta-1}%
\end{align}
Here, we have introduced the following notations for the momenta depending
coefficients in an arbitrary realization%
\begin{align}
\Omega_{\mu}^{\pm}(k)  &  =\pm\frac{i}{2}\left(  2c-\frac{1}{2}\right)
\mathbf{A}_{\pm}k_{\mu},\nonumber\\
\Lambda_{\pm\mu}^{\nu\sigma}(k)  &  =\pm\frac{i}{2}\left(  4c-\frac{1}%
{2}\right)  \mathbf{A}_{\pm}\frac{k_{\mu}k^{\nu}k^{\sigma}}{k^{2}%
},\label{not-2}\\
\Gamma_{\pm\mu}^{\omega\nu\sigma}(k)  &  =\left(  2c-\frac{1}{2}\right)
\mathbf{A}_{\pm}\eta^{\omega\nu}\frac{k_{\mu}k^{\sigma}}{k^{2}},
\label{not-3}\\
\Upsilon_{\pm\mu}^{\omega\sigma\nu}(k)  &  =\left(  4c-\frac{1}{2}\right)
\mathbf{A}_{\pm}\delta^{\omega\sigma}\frac{k_{\mu}k^{\nu}}{k^{2}}.
\label{not-4}%
\end{align}
We note that while $j_{\pm}^{\mu}$ -, $\alpha^{\pm}$ - and $\beta^{\pm}$ -
waves, respectively, are common to all noncommutative fluids from the class
described by the linearized action (\ref{S-lin}), the potentials $\theta^{\pm
}(x)$ depend on the details of each particular model. As was mentioned in the
previous sections, a model can be specified by choosing the arbitrary function
$f(\rho_{0})$ and the fluid density function $\rho_{0}(x)$.

A simple class of models is given by $f(\rho_{0})=\lambda\rho_{0}^{2}/2$ where
$\lambda$ is a real parameter and $\rho_{0}$ is an arbitrary function. This
type of commutative fluids has been discussed in \cite{Nyawelo:2003bv} in the
K\"{a}hler parametrization, while in \cite{Holender:2008qj} it was shown that
they can be quantized using canonical methods. Then the equation
(\ref{Sol-theta-1}) takes the following simpler form%
\begin{equation}
\left(  \partial^{2}-\frac{2}{5s}\right)  \partial_{\mu}\theta^{\pm}%
=-\frac{2i}{5s}\mathbf{D}_{\pm}k_{\mu}, \label{Sol-theta-2}%
\end{equation}
The above equation can be integrated by standard methods or the solutions can
be simply guessed. In either way, we find that the solutions $\theta^{\pm}$
that correspond to $j_{\pm}^{\mu}$ -, $\alpha^{\pm}$ - and $\beta^{\pm}$-
waves are linear functions on $x$%
\begin{equation}
\theta^{\pm}(x)=\mathbf{\vartheta}^{\pm}\pm i\mathbf{D}_{\pm}\left\langle
kx\right\rangle , \label{Sol-theta-3}%
\end{equation}
where $\mathbf{\vartheta}^{\pm}$ are arbitrary integration constants. The fact
that the $\theta^{\pm}$ - potentials are linear implies that they contribute
by constant pieces to the fluid velocity $v_{\mu}^{\pm}\sim$ $\partial_{\mu
}\theta^{\pm}+\cdots$ in the commutative limit.

The interpretation of the above solutions is that of the noncommutatively
deformed monocromatic waves propagating in the relativistic perfect fluid with
a simple geometry of fields determined by the orthogonality of the gradients
of the fluid potentials on the density current. These waves are solutions to
the dynamics of the long wavelength degrees of freedom of a system on the
noncommutative space-time which reduces in the commutative limit to the
relativistic perfect fluid.

\subsection{The energy-momentum tensor}

An important quantity that contains information about the thermodynamics of
the relativistic fluid is the energy-momentum tensor. For the effective field
theory that describes the noncommutative fluid, it has the general form given
by the equation (\ref{A-2}). By expanding in powers of $s$ and retaining only
the linear terms we obtain the linearized energy-momentum tensor $T_{\nu}%
^{\mu}(j,\theta,\alpha,\beta)$. From it, one can calculate the energy-momentum
tensor of the wave potentials $T_{\nu}^{\mu}(j_{\pm},\theta^{\pm},\alpha^{\pm
},\beta^{\pm})$. After lenghty computations, one can show that the
energy-momentum tensor takes the following from%
\begin{align}
T_{\nu}^{\mu}(j^{\pm},\theta^{\pm},\alpha^{\pm},\beta^{\pm})  &
=\mathbf{A}_{\pm}\left(  \mathbf{D}_{\pm}-\mathbf{C}^{\pm}\mathbf{B}^{\pm
}\right)  e^{\pm i\left\langle kx\right\rangle }\left(  \eta_{\nu}^{\mu}%
-\frac{k^{\mu}k_{\nu}}{k^{2}}\right) \nonumber\\
&  +\frac{s\lambda}{4}\left(  \mathbf{A}_{\pm}\right)  ^{2}\frac
{e^{\pm2i\left\langle kx\right\rangle }}{k^{2}}\left[  \left(  \frac{7}{2}%
\pm2i\left\langle kx\right\rangle \right)  k^{\mu}k_{\nu}\pm2ik^{2}x^{\mu
}k_{\nu}\right] \nonumber\\
&  +\frac{s}{2}\mathbf{A}_{\pm}\mathbf{D}_{\pm}e^{\pm i\left\langle
kx\right\rangle }\left[  \left(  1\mp i\left\langle kx\right\rangle \right)
k^{\mu}k_{\nu}\pm ik^{2}\left\langle kx\right\rangle \eta_{\nu}^{\mu}\right]
\nonumber\\
&  \mp\frac{is}{2}\mathbf{A}_{\pm}\mathbf{B}^{\pm}\mathbf{C}^{\pm}e^{\pm
i\left\langle kx\right\rangle }\left[  -\left(  8\mp7i\right)  \left\langle
kx\right\rangle k^{\mu}k_{\nu}+\left(  3\pm5i\right)  k^{\mu}k_{\nu}+\left(
2\mp i\right)  k^{2}x^{\mu}k_{\nu}\right]  . \label{em-ten-waves}%
\end{align}
Note that the energy-momentum tensor of the single mode wave fluids in the
noncommutative fluid model studied above is complex as it is in the case of
spinor fields and $N=1$ chiral supergravity in the commutative space-time. As
can be seen from the equation (\ref{em-ten-waves}), the imaginary part is a
consequence of the noncommutativity of space-time. The noncommutative
deformation of the energy-momentum tensor of the relativistic fluid contains
off-diagonal terms that are not proportional to the fluid velocity component
as is the case in the commutative space-time, but rather with the space-time
coordinates. This shows that as the wave moves in the space, there is an
exchange of energy and momentum of the different terms from
(\ref{em-ten-waves}) with the noncommutative space-time. A similar phenomenon
occurs with nonlocal systems in the graviational field where the normal modes
exchange energy with the gravitational field viewed as a fixed (non-trivial)
background which can lead to thermalization.

\section{Concluding remarks}

In this paper we have addressed the question of the dynamics of the
noncommutative fluid in the Snyder space at first order in the expansion of
the star-product, co-product and the anti-pode in powers of the noncommutative
parameter $s$ and have shown that the linearized equations of motion of the
fluid density and the fluid potentials form a system of linear partial
differential equations. We have determined a class of perturbative analytic
solutions of these equations that describe monocromatic plane waves of
$j^{\mu}$ and $\alpha$ and $\beta$ potentials. For the class of fluids
characterized by $f(\rho_{0})=\lambda\rho_{0}^{2}/2$ the equation of motion of
the potential $\theta$ admits a linear solution. We have calculated the
energy-momentum tensor of these solutions and shown that it has a complex
structure determined by the noncommutativity of space-time at first order in
$s$. The physical interpretation of the diagonal terms that are proportional
to $s$ is that they are proportional to the noncommutative corrections to the
energy density and pressure density of the monocromatic waves propagating in
the relativistic perfect fluid. However, there are non-zero off-diagonal terms
which are not proportional with the velocity of the fluid particles and are
not present in the perfect commutative fluid. These terms arise at the long
wave scale from the gradients of the fluid potentials and density current
along the noncommutative directions of space-time. From the form of the
energy-momentum tensor we conclude that its components proportional to $s$
describe the exachange of energy and momentum between the monocromatic waves
and the noncommutative space-time similarly to the phenomenon that occurs in
the gravitational field.

We note that the noncommutative parameter determines at first order the
dispersion relations of the monocromatic wave potentials as follows%
\begin{equation}
k_{j}^{2}-\frac{2}{s}=0,\qquad k_{\alpha}^{2}=\frac{1-s}{3s},\qquad k_{\beta
}^{2}=0,\label{disp-rel}%
\end{equation}
which can be used to determine the on-shell solutions and energy-momentum
tensor. In the commutative limit $s\rightarrow0$ the dispersion relations for
the density of current density and $\alpha$ - potential diverge as do all
terms analogous to the masses in the equations of motion. This shows that the
noncommutative waves behave as infinitely massive fields in the commutative
limit. Their dynamics in this case is no longer described by wave - equations
since the formal step taken to derive these equations is invalid in this
limit. The proper equations of motion of the commutative fluid are defined by
the commutative limit of the equations (\ref{eq-mot-j}) - (\ref{eq-mot-beta}).
This render, for example, a conservation equation for the current density.
Another important property of the solutions obtained in (\ref{Sol-j-1}),
(\ref{Sol-beta-1}), (\ref{Sol-alpha-1}) and (\ref{Sol-theta-3}) is that the
geometry of the waves is such that the gradients of $\alpha$ and $\beta$
potentials is normal to the direction of $j^{\mu}$. However, the plane waves
do not create vorticity in the linear approximation of the noncommutative
fluid regardless the choice of the parameters $f$ and $\rho_{0}$ which is an
expected result since the vorticity is not a well defined concept in the
Snyder space. Nevertheless, the model can still produce in the commutative
limit $s\rightarrow0$ a rotational fluid.

The results presented in this paper provide a new insight in the dynamics of
the noncommutative fluids. They are interesting by themselves, as well as for
offering a novel explicit construction of noncommutative effective field
theories in which the energy-momentum tensor is calculated from the scratch in
the linear approximation. It is interesting to study further the possible
solutions of the noncommutative fluid dynamics and their thermodynamics as
well as to extend it to incorporate charges, which could be useful for
developping phenomenological models. We are going to report on these aspects
in future.

\textbf{Acknowledgements} The work of M. C. B. A. was partially supported bt
the CNPq Grant: 306276/2009-7. I. V. V. would like to thank to N. Berkovits
for hospitality at ICTP-SAIFR where this work was accomplished and to H.
Nastase and A. Mikhailov for useful discussions. Also, the authors would like
to acknowledge the anonimous referee for constructive comments.

\end{document}